\documentclass[10pt,conference]{IEEEtran}

\usepackage[pdftex]{graphicx}
\usepackage{amsmath}
\DeclareMathSizes{10}{10}{8}{7}   

\begin{document}

\title{Random Linear Network Coding for Time-Division Duplexing: Queueing Analysis}

\author{
\authorblockN{Daniel E. Lucani}
\authorblockA{RLE, MIT\\
Cambridge, Massachusetts, 02139 \\
Email: dlucani@mit.edu}
\and
\authorblockN{Muriel M\'edard}
\authorblockA{RLE, MIT\\
Cambridge, Massachusetts, 02139\\
Email: medard@mit.edu}
\and
\authorblockN{Milica Stojanovic
}
\authorblockA{Northeastern University \\
Boston, Massachusetts, 02115\\
Email: millitsa@mit.edu
}
}
%

\maketitle

\begin{abstract}
We study the performance of random linear network coding for time division duplexing channels with Poisson arrivals. We
model the system as a bulk-service queue with variable bulk size. 
A full characterization for random linear network coding is provided for time division duplexing channels \cite{lucaniInfocom09} by means of the moment generating function.
We present numerical results for the mean number of packets in the queue and consider the effect of the range of allowable bulk sizes. We show that there exists an optimal choice of this range that minimizes the mean number of data packets in the queue. 
\end{abstract}

\section{Introduction}

Reference \cite{lucaniInfocom09} considered the use of network coding in channels in which time division duplexing is necessary, i.e. a node can only transmit or receive, but not both at the same time. This type of channel is usually called half-duplex, but the term time division duplexing (TDD) was used to emphasize that the channel is not used in any pre-determined fashion, but instead may {\em vary} the amount of time allocated to transmit and receive.  

In particular, Reference \cite{lucaniInfocom09} studied the problem of transmitting $M$ data packets through a link using random linear network coding with the objective of minimizing the expected time to complete transmission of the $M$ data packets. Reference \cite{lucaniICC09} focused on the problem of energy consumption of this scheme showing that there exists, under the minimum energy criterion, an optimal number of coded data packets to be transmitted back-to-back before stopping to wait for an acknowledgment (ACK). 

The assumption in References \cite{lucaniInfocom09} and \cite{lucaniICC09} was that the source had $M$ data packets in its buffer before starting transmission. In a more realistic network setting, this buffer may sometimes empty or contain fewer than $M$ packets awaiting transmission. Then, the source node must choose to either wait for additional packets
to arrive, or take those packets in the buffer and start performing random linear coding. 

The problem of queueing for network coding systems has been considered previously to account for burstiness or losses. Some of this work considered the case in which feedback is available, e.g. \cite{Sundararajan07}. References \cite{shrader06} and \cite{shrader07} studied a system with random linear coding, slotted time, and a Bernoulli arrival process. However, previous work has not considered timing or TDD constraints. Our work considers the problem in which the channel is TDD, where the time is not slotted and, more importantly, the service time depends on the size of the bulk.

We study these scenarios for random linear network coding (RLNC) for TDD channels when the data packets arrive randomly at the source node according to a Poisson process. This problem can be modeled as a bulk service queue with a general service process. Bulk or batch service queues have been studied widely, e.g. \cite{chaudhry83}. However, 
the service time for RLNC TDD will depend on the transmission time of both the coded packets and ACK packets, on the number of coded packets that are sent, and on the propagation time, as shown in \cite{lucaniInfocom09}. This means that we have a bulk queue with a general service time, where the service time depends on the size of the bulk, which is not common in the existing studies. Reference \cite{barlev07} recently studied this problem with Poisson arrivals, calling it the $M/G^{(m,K)}/1$ queue, where the size of the bulk can range between $m$ and $K$. We build on this work to develop the queueing model of RLNC TDD.  

We provide numerical results for the mean queue size of the system for different choices of the arrival rate, and the pair $(m,K)$. We show that the choice of the pair $(m,K)$ can greatly impact the mean queue size of the system, suggesting that there is an optimal choice that minimizes the mean queue size. We show that fixing $m =1$ and choosing $K$ according to the arrival rate $\lambda$ is optimal in this sense. Also, we show that having a fixed batch size, i.e. $m=K$, is not the optimal configuration in general.


	 The paper is organized as follows. In Section 2, we describe the system model and provide an expression for the moment generating function of the service time. In Section 3, we present the queueing model. In Section 4, numerical examples are provided. Conclusions are summarized in Section 5.

\section{System Model}

\begin{figure}[t]
\centering	
\includegraphics[height=1.0in,width=3.5in,keepaspectratio]{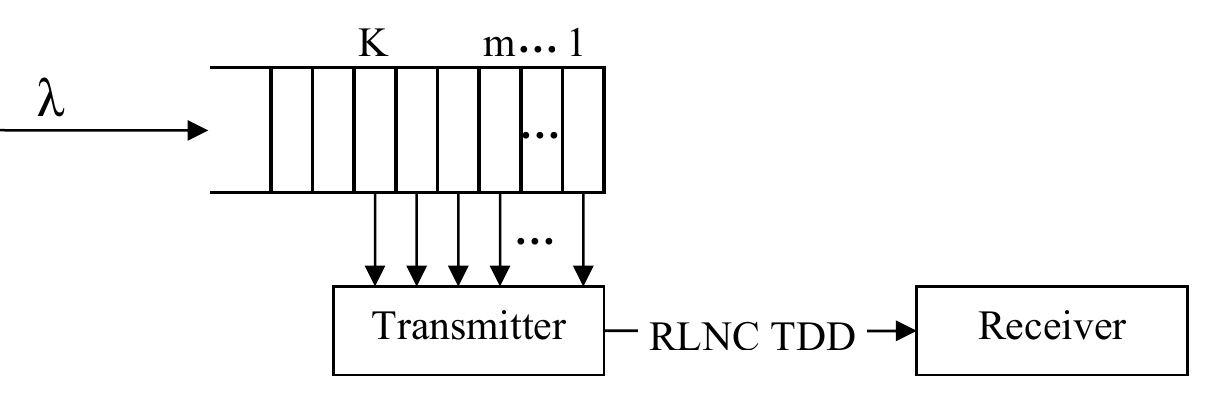}
\caption{Queue model studied in this work.}
\label{QueueModel.tag}
\end{figure}

We consider each data packet to be of fixed-length, arriving to a source node through a Poisson process with rate $\lambda $ packets/s. Upon
arrival, the data packet is placed in a buffer to await encoding and transmission to the receiver, as in Figure~\ref{QueueModel.tag}. 
The buffer forms a first-in-first-out (FIFO) queue. The service time of the queue is given by the time it takes to transmit a group of $M$ packets taken from the queue using random linear network coding for TDD channels \cite{lucaniInfocom09}.  The size of the group of packets is variable, where $m \leq M \leq K$. The pair $(m,K)$ constitutes the range of the bulk size or number of packets taken to perform random linear network coding \cite{ho06}. If the buffer has fewer than $m$ data packets, the system will wait until $m$ packets arrive before providing service. If the buffer contains more than $K$ packets, the system will service exactly $K$ packets. Finally, if the buffer has $M$ packets with $m \leq M \leq K$, then the system will service $M$ packets. Note that the service time depends on the number of data packets taken from the queue at any time, i.e. the service time distribution is general but it depends on the size of the batch being transmitted. Thus, we can use the bulk queueing model $M/G^{(m,K)}/1$ developed in \cite{barlev07} to study the problem.

\begin{figure}[t]
\centering	
\includegraphics[height=1in,width=3.5in, keepaspectratio]{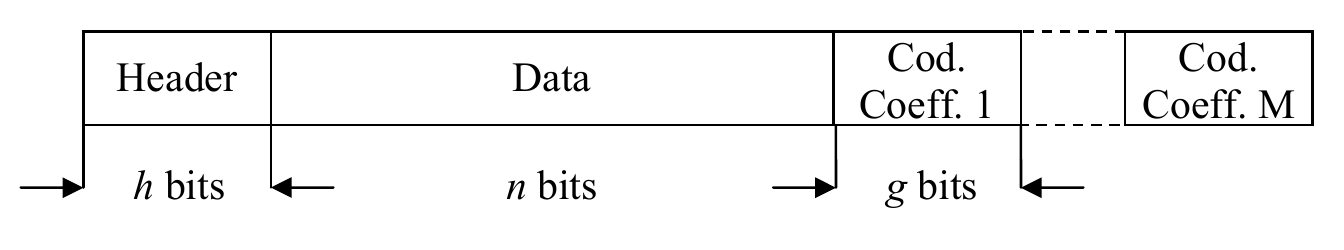}
\caption{\cite{lucaniInfocom09} Structure of coded data packet: a header of size $h$ bits, $n$ data bits, $M$ coding coefficients of size $g$~bits each.}
\label{PacketFrame.tag}
\end{figure}

	We consider that once the sender in a link has $M$ data packets, it wants to transmit them reliably at a given link data rate $R$ [bps]. The channel is modeled as a packet erasure channel. Nodes can only transmit or receive, but not both at the same time. The sender uses random linear network coding to generate coded data packets, which means that each coded data packet contains a linear combination of the $M$ data packets of $n$ bits each, as well as the random coding coefficients used in the linear combination. Each coefficient is represented by $g$ bits. For encoding over a field size $q$, we have that $g = \log_2 q$ bits. A coded packet is preceded by an information header of size $h$~bits. Thus, the total number of bits per packet is $h + n + gM$. Figure \ref{PacketFrame.tag} shows the structure of each coded packet. 

	As in \cite{lucaniInfocom09}, the sender can transmit coded packets back-to-back before stopping to wait for the ACK packet. The ACK packet feeds back the number of degrees of freedom (dof), that are still required to decode successfully the $M$ data packets. Transmission begins with the $M$ information packets taken from the queue, which are encoded into $N_M \geq M$ random linear coded packets, and transmitted. If all $M$ packets are decoded successfully, the process is completed and a new batch of data packets can be serviced. Otherwise, the ACK informs the transmitter how many dofs are missing, say $i$. The transmitter then sends $N_i$ coded packets, and so on, until all $M$ packets have been decoded successfully. 

\begin{figure}[t]
\centering	
\includegraphics[height=1in,width=3.5in]{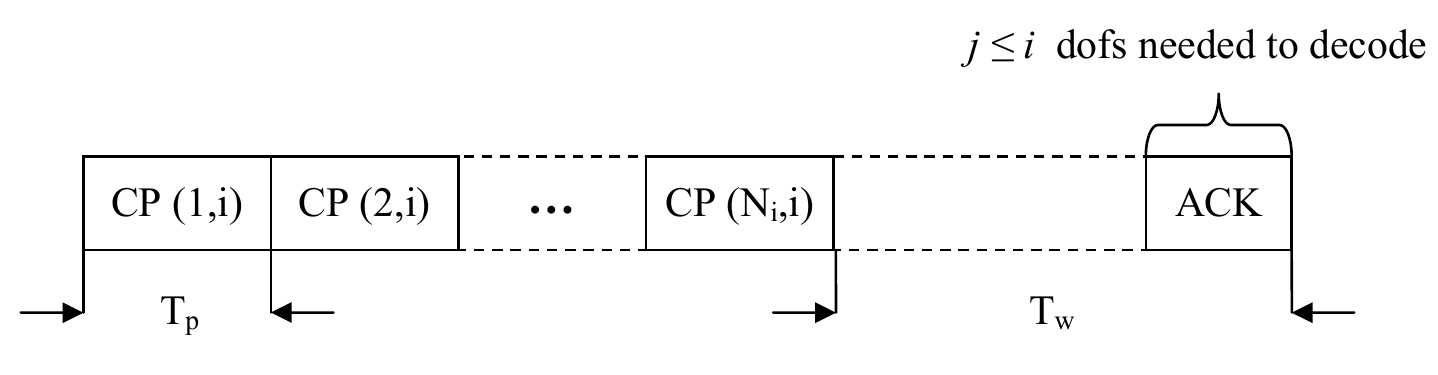}
\caption{\cite{lucaniInfocom09} Network coding TDD scheme.}
\label{Protocol.tag}
\end{figure}

Figure~\ref{Protocol.tag}, illustrates the time window allocated to the system to transmit $N_i$ coded packets. Each coded packet $CP(1,i)$, $CP(2,i)$, etc. is of duration $T_p$. The waiting time $T_w$ is chosen so as to accommodate the propagation delay and time to receive and ACK.  
The problem of choosing the optimal number $N_{i}$ of coded packets to be transmitted back-to-back when $i$ dofs are required at the receiver in order to decode the information was studied in \cite{lucaniInfocom09}.  


	This process can be modelled as a Markov chain. A state is defined as the number of dofs required at the receiver to decode successfully the $M$ packets. Thus, the states range from $M$ to 0. This is a Markov chain with $M$ transient states and one recurrent state (state 0). Note that the time and energy spent in each state depends on the state itself, because $N_{i} \neq N_{j}, \forall i \neq j$ in general. The transition probabilities from state $i$ to state $j$ ($P_{i\rightarrow j}$) are defined in \cite{lucaniInfocom09}.
	
\subsection{Moment Generating Function}
Let us define the moment generating function of the completion time when the Markov Chain starts at state $n$ as 
\begin{equation}
M_{T,n} (s) = \sum_t \exp(s t ) P_{T}(T = t)
\end{equation}
where $P_{T}(T = t)$ is the probability of the completion time being $t$. Note that $M_{T,n}(s)$ is the moment generating function of the completion time when $n$ data packets are taken by the source to be transmitted reliably to the receiver. 

Using the Markov Chain structure of the problem, it can be shown that $M_{T,n}(s)$ can be re-stated as
\begin{equation} \label{MomentGenerating2}
M_{T,n} (s) = \sum_{m_n \geq 1} \sum_{m_{n-1} \geq 0} \cdot \cdot \cdot \sum_{m_1 \geq 0} \exp\left( s \sum_{i=1}^n m_i T^i \right) C_n A_n
\end{equation}
where $T^i$ is the deterministic time required to send $N_i$ coded packets and wait for an ACK when the Markov chain is in state $i$, i.e. $T^i = N_i T_p + T_w$. The constant $C_n$ captures the effect of returning to the same state repeatedly, while $A_n$ captures the different paths that can be traversed without repetition of a state.

The expression for $C_n$ is
\begin{equation}
C_n = \prod_{j=1}^{n} {P_{j\rightarrow j}}^{m_j -1}. \notag
\end{equation}

The coefficient for $A_n$ can be shown to obey a recursive expression of the form
\begin{equation} \label{AnFormula}
A_n = \mathbf{1}_{ \{  m_n > 0 \} } \left[   \sum_{j=0}^{n-1} P_{n\rightarrow j} \left( \prod_{i=j+1}^{n-1} P_{i\rightarrow i} \mathbf{1}_{ \{ m_i = 0 \} } \right) A_j   \right] \notag
\end{equation}
with $A_1 = P_{1\rightarrow0} \mathbf{1}_{ \{ m_1 > 0 \} }$. The indicator function $\mathbf{1}_{ \{ s \in S \} }$ is 1 when $s \in S $ and zero otherwise.

Substituting expression~\eqref{AnFormula} into \eqref{MomentGenerating2} we obtain the following recursive equation for the moment generating function
\begin{equation} \label{MomentGeneratingFunction}
M_{T,n}(s) = \frac{\exp(s T^n)}{1 - P_{n\rightarrow n}\exp(s T^n)} \sum_{i = 0}^{n-1} P_{n\rightarrow i}M_{T,i}(s)
\end{equation}
with $M_{T,0}(s) = 1$. 

Finally, note that the same structure is valid for computing the energy needed to complete transmission. To do so, one would substitute $T^i$ by $E^i$, and $M_{T,n}(s)$ by $M_{E,n}(s)$.

\section{Queueing Model}
 The system model discussed in the previous section corresponds to the bulk queueing model $M/G^{(m,K)}/1$ studied in \cite{barlev07}. This bulk queueing model considers Poisson arrivals and a general service time that depends on the bulk size.   

The transition probability of the number of packets in the queue is given by \cite{barlev07}
\begin{eqnarray}
P =
\left [ 
\begin{array}{cccccccc} 
a_0^{(m)}&a_1^{(m)}& \cdot \cdot \cdot & a_K^{(m)} & a_{K+1}^{(m)}& \cdot \cdot \cdot\\
a_0^{(m)}&a_1^{(m)}& \cdot \cdot \cdot & a_K^{(m)} & a_{K+1}^{(m)}& \cdot \cdot \cdot\\
:&:&:&:&:&:\\
a_0^{(m)}&a_1^{(m)}& \cdot \cdot \cdot & a_K^{(m)} & a_{K+1}^{(m)}& \cdot \cdot \cdot\\
a_0^{(m+1)}&a_1^{(m+1)}& \cdot \cdot \cdot & a_K^{(m+1)} & a_{K+1}^{(m+1)}& \cdot \cdot \cdot\\
:&:&:&:&:&:\\
a_0^{(K)}&a_1^{(K)}& \cdot \cdot \cdot & a_K^{(K)} & a_{K+1}^{(K)}& \cdot \cdot \cdot\\
0&a_0^{(K)}& \cdot \cdot \cdot & a_{K-1}^{(K)} & a_{K}^{(K)}& \cdot \cdot \cdot\\
0&0&a_0^{(K)}& \cdot \cdot \cdot & a_{K-2}^{(K)} & \cdot \cdot \cdot\\
:&:&:&:&:&:\\
\end{array}  
\right] \notag
\end{eqnarray}
where $a_k^{(j)}$ is the probability of $k$ arrivals during a service of type $j$. 

Let us define
\begin{equation}
A^{(j)}(z) = \sum_{k=0}^{\infty } a_k^{(j)}z^k .
\end{equation}

We can use a similar analysis as that of Reference \cite{barlev07} to prove that
\begin{equation}
A^{(j)}(z) = M_{T,j}(\lambda (z - 1))
\end{equation}
and that
\begin{equation}
a_k^{(j)} = \frac{1}{k!} \frac{\partial ^k }{\partial z^k} M_{T,j}(\lambda (z - 1)) \Big | _{z=0}.
\end{equation}

The system is stable if and only if $\lambda < K \mu_K$, where $1/\mu_K$ is the mean service time when the bulk size is $M = K$, where $1/\mu_j = \frac{\partial }{\partial z} M_{T,j}(z) \Big |_{z = 0}$.

Let us denote by $\Pi (z) = \sum_{i=0}^{\infty} \pi_i z^i$ the corresponding generating function of the stationary probabilities. Reference \cite{barlev07} showed that $\Pi(z)$ can be expressed as
\begin{eqnarray}
\Pi(z) &= \frac{A^{(K)}(z) \sum_{i=0}^K \pi_i z^i - z^KA^{(m)}(z)\sum_{i=0}^m \pi_i}{A^{(K)}(z) - z^K} \notag \\
& - \frac{\sum_{i=m+1}^K \pi_i A^{(i)}(z)}{A^{(K)}(z) - z^K} 
\label{StationaryProb}
\end{eqnarray}
which provides an expression for $\Pi(z)$ in terms of its first $K+1$ coefficients $\pi_0,...,\pi_K$. Determining these $K+1$ coefficients provides a full characterization of the stationary probabilities \cite{barlev07}. 
Reference \cite{barlev07} proves that $A^{(K)}(z) - z^K$ has exactly $K$ zeros satisfying $|z|\leq 1$ assuming that $A^{(K)}(z)$ has a radius of convergence greater than one. Denoting the roots as $1, z_1,...,z_{K-1}$ and assuming that they are different, note that the numerator of \eqref{StationaryProb} has to vanish for $z_1,...,z_{K-1}$ which gives us $K-1$ linear equations 
\begin{eqnarray}
&A^{(K)}(z_k) \sum_{i=0}^K \pi_i z_k^i - z_k^KA^{(m)}(z_k)\sum_{i=0}^m \pi_i \notag\\& - \sum_{i=m+1}^K \pi_i A^{(i)}(z_k) =0
\end{eqnarray}
for $k = 1,...K-1$.
Also, the numerator vanishes trivially for $z = 1$ for both the numerator and the denominator in~\eqref{StationaryProb}. We thus need one more linear equation. To obtain this we use l'H\^ ospital's rule to exploit the fact that $\Pi(1) = 1$. This translates to 
\begin{eqnarray}
1 = \sum_{i=0}^{m} \left[ 1 + \frac{i - \lambda /\mu_m     }{\lambda /\mu_K - K } \right] \pi_i + \sum_{i=m+1}^{K} \left[ \frac{ \lambda / \mu_K + i - \lambda / \mu_i }{\lambda / \mu_K - K} \right] \pi_i \notag
\end{eqnarray}
where we have used the fact that $\frac{\partial A^{(i)}(z)}{\partial z} \Big |_{z=1} = \lambda / \mu_i$. 
Since, $\lambda / \mu_K \neq K$ in general, the denominator will not become zero as $z \rightarrow 1$. In fact, if the system is stable this condition will be satisfied.    

The final linear equation to fully characterize $\Pi(z)$ given in Reference \cite{barlev07} is
\begin{equation}
(a_0^{m} - 1) \pi_0 +  a_0^{m} \pi_1 + \cdot \cdot \cdot +   a_0^{m} \pi_m +   a_0^{m+1} \pi_{m+1} +   a_0^{K} \pi_K =0
\end{equation}

\subsection{Queue of Finite Capacity}

The general solution requires the calculation of the roots of $A^{(K)}(z) - z^K$, which can result in
numerical inaccuracies in practice because $A^{(K)}$ has exponential terms. Also, calculating the roots is increasingly difficult when the decision variable $K$ assumes a larger value. For these reasons, we simplify the problem considering that the system has a capacity of $B$ packets waiting to be serviced, i.e. without considering those that are being transmitted. 
The transition probability for this case is
\begin{eqnarray}
P = 
\left [ 
\begin{array}{ccccccc} 
a_0^{(m)}&a_1^{(m)}& \cdot \cdot \cdot & a_{B-1}^{(m)} & R{(B-1,m)}\\
a_0^{(m)}&a_1^{(m)}& \cdot \cdot \cdot & a_{B-1}^{(m)} & R{(B-1,m)}\\
:&:&:&:&:\\
a_0^{(m)}&a_1^{(m)}& \cdot \cdot \cdot & a_{B-1}^{(m)} & R{(B-1,m)}\\
a_0^{(m+1)}&a_1^{(m+1)}& \cdot \cdot \cdot & a_{B-1}^{(m+1)} & R{(B-1,m+1)}\\
:&:&:&:&:\\
a_0^{(K)}&a_1^{(K)}& \cdot \cdot \cdot & a_{B-1}^{(K)} & R{(B-1,K)}\\
0&a_0^{(K)}& \cdot \cdot \cdot & a_{B-2}^{(K)} & R{(B-2,K)}\\
:&:&:&:&:\\
\cdot \cdot \cdot & a_0^{(K)} & \cdot \cdot \cdot & a_{B-K}^{(K)} &R{(B-K,K)}\\
\end{array}  
\right] \notag
\end{eqnarray}
where $R{(k,l)} = 1 - \sum_{j=0}^{k} a_{j}^{(l)}$.

In order to compute the stationary distribution, it suffices to solve $\bar \pi = P \bar \pi$, with $\bar \pi = {[\pi_0, \pi_1, ..., \pi_B ]}^T$, under the constraint that $\sum_{i=0}^{B} \pi_i = 1$.

\subsection{Performance Analysis}

We will consider two metrics in order to study performance of the system. 
First, the mean queue size defined as
\begin{equation}
E[Q] = \sum _{i=0}^{B} i \pi_i
\end{equation}
for the case in which the queue capacity is $B$. If there is no constraint on the capacity, we simply let $B \rightarrow \infty $. 

The second metric is the mean batch size in steady state, which can take values $\{m, m+1, ..., K \}$. Defining $Z_{(m,K)}$ as the batch size for a choice of $(m,K)$, then
\begin{equation}
E[Z_{(m,K)}] = m \sum _{i=0}^{m} \pi_i + \sum _{i=m+1}^{K-1} i \pi_i + K \sum _{i=K}^{B}\pi_i,
\end{equation}
where $\sum _{i=K}^{B} \pi_i = 1 - \sum _{i=0}^{K-1}\pi_i$, and if there is no constraint on capacity, again we let $B \rightarrow \infty $.

\begin{figure}[t]
\centering	
\includegraphics[height=3.4in,width=3.4in,keepaspectratio]{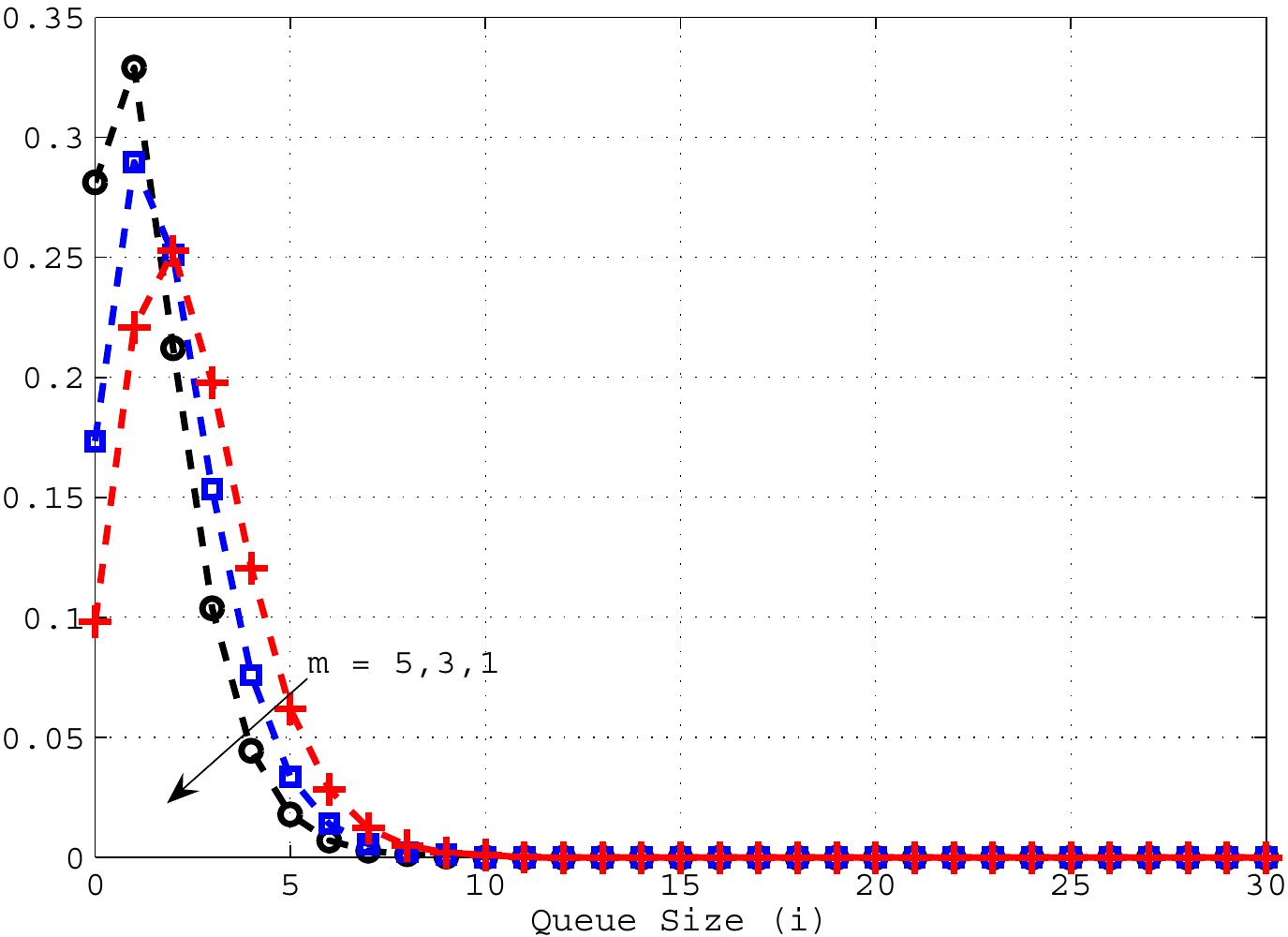}
\caption{Stationary distribution $\pi_i, i = 0, 1,...,B$ for $\lambda = 30$~packets/s, $K = 5$, $B = 30$, and different values of $m$.}
\label{StationaryProb.tag}
\end{figure}

\section{Numerical Results}

This section provides numerical examples that show the performance of our network coding scheme for different settings of $(m,K)$ and arrival rate $\lambda$. We use the mean queue size defined in the previous section as our metric of interest. We use a high latency channel with packet erasure probability $Pe = 0.2$, propagation time of $12.5$~ms, data packets of $10,000$~bits, $g = 20$~bits, a rate $R = 1.5$~Mbps, a header of $h= 80$~bits, and the ACK packet has $100$~bits. We assume that the number of coded packets to be sent back-to-back ($N_i$) are chosen to minimize the mean transmission time as in \cite{lucaniInfocom09}.  

Figure~\ref{StationaryProb.tag} shows the stationary distribution for $\lambda = 30$~packets/s, $ K = 5$, $B = 30$ and different values of $m$. This figure shows that the high probability states correspond to small number of packets in the queue. The probability of large queue sizes when the system operates in steady state is very low. Figure~\ref{StationaryProb.tag} also shows that for low values of $m$, the stationary distribution is concentrated in the lower values of the queue size $i$. As $m$ increases, the stationary distribution spreads over larger values of the queue size $i$. 
 
Table~\ref{Table_lowlambda_25ms} shows the mean queue size when $\lambda = 1$~packet/s under different configurations of the pair $(m,K)$, with $m \leq K$. We observe that the mean queue size shows greater dependence on the value of $m$ than on the value of $K$. For example, increasing the value of $K$ when $m=1$ shows little variation in the mean queue size, while increasing $m$ with any value of $K>1$ increases the mean queue size. In terms of minimizing the mean queue size for low values of $\lambda$, this means that we should allow transmission of bulks of size 1. This is the case because waiting for additional packets before transmitting is costly, considering that the time between packet arrivals might be larger than the mean service time for a single data packet.   
For the case of bulks of size 1, our scheme transmits several copies of the packet back-to-back before stopping for an ACK, which is similar to the idea presented in \cite{sastry75}. 

Table~\ref{Table_highlambda_25ms} shows the mean queue size when $\lambda = 30$~packet/s under different configurations of the pair $(m,K)$, with $m \leq K$. For this $\lambda$, we do not consider the setting $m=K=1$ because it is not stable for the infinite capacity case ($B \rightarrow \infty $) and will present very high packet drops when the capacity $B$ is finite. 
Again, there is an advantage of allowing $m=1$ for the studied cases, in terms of reducing the mean queue size. 

Tables~\ref{Table_lowlambda_25ms} and ~\ref{Table_highlambda_25ms} also show the mean batch size for $\lambda = 1$~packet/s and $\lambda = 30$~packet/s, respectively. The mean batch size is biased by the value of $m$ and $K$. However, it gives us some intuition about the operation of the system. For example, Table~\ref{Table_lowlambda_25ms} shows that when $m$ is too large with respect to the arrival rate, the mean batch size is close to $m$. This means that on the average the system services the batches much faster than the time it takes the queue to fill with $m$ new packets. Thus, the system will be idle for long periods of time just waiting for the queue to fill to the required $m$. Only with small probability the batches will contain more than $m$ packets. Of course, if $m=K$ the batch size will always be $m$ as seen in the tables.  

\begin{figure}[t]
\centering	
\includegraphics[height=3in,width=3.2in,keepaspectratio]{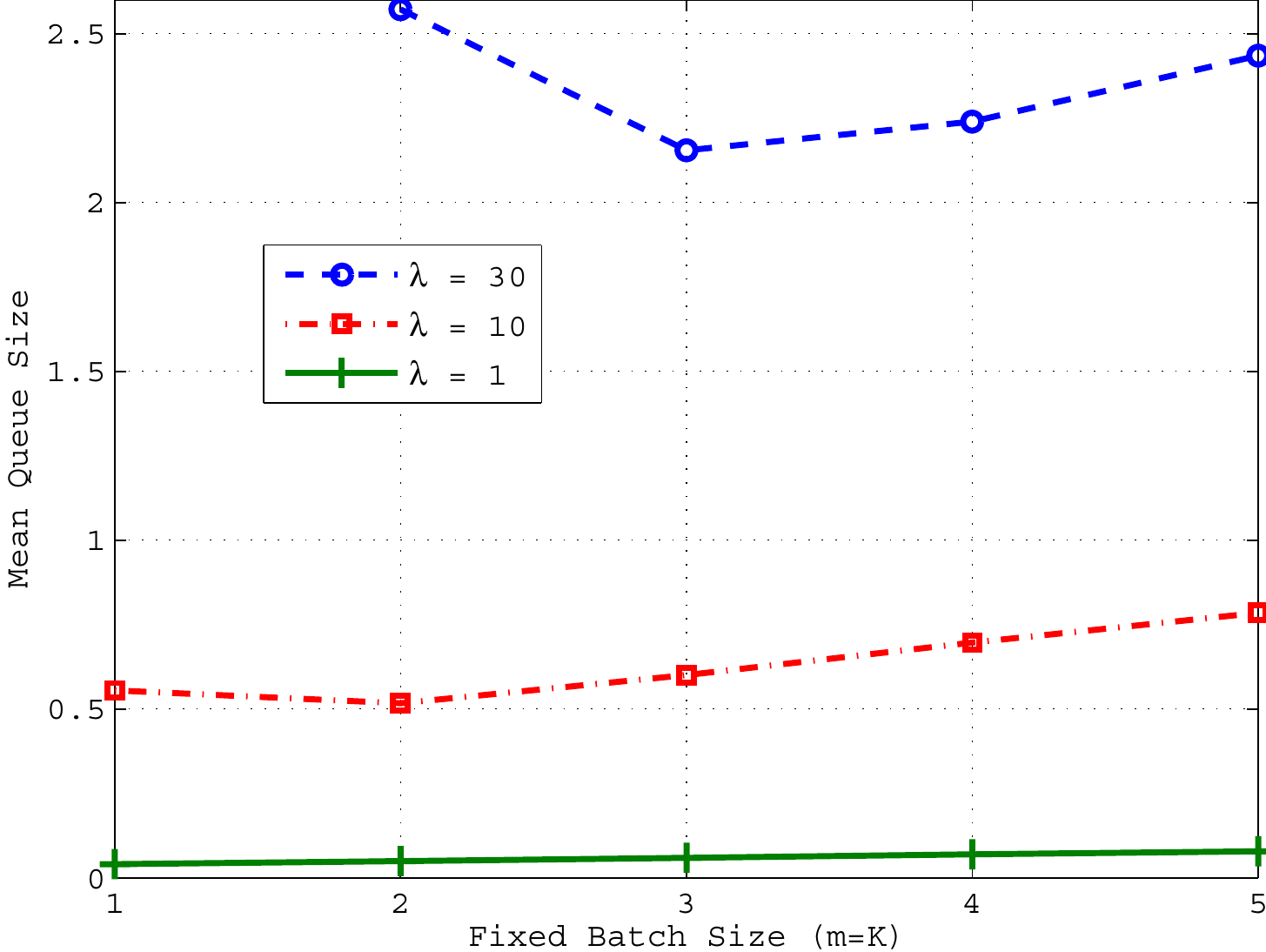}
\caption{Mean queue size for the fixed batch size case ($m=K$) with $B = 30$.}
\label{Queuesize.tag}
\end{figure}

Let us consider the case of a fixed batch size, i.e. $m=K$. Tables~\ref{Table_lowlambda_25ms}, \ref{Table_highlambda_25ms} and Figure~\ref{Queuesize.tag} show that the optimal choices are $m=K=1$, $m=K=2$, and $m=K=3$ for $\lambda = 1$~packet/s, $\lambda = 10$~packet/s, and $\lambda = 30$~packet/s, respectively. 
However, we notice that the mean queue size is larger than other configurations of $(m,K)$ without the fixed batch size restriction. For example, the optimal fixed batch size configuration  for $\lambda = 30$~packet/s is $53$~\% larger than the $(m,K) = (1,5)$ configuration presented in the table.
Thus, we observe that having a fixed batch size is not the optimal configuration in general.

\begin{table}
\caption{Mean Queue Size for different $(m,K)$ configurations. The parameters used are $\lambda = 1$~packet/s}
\centering
\label{Table_lowlambda_25ms}
\begin{tabular}{|c||c|c|c|c|c|}
\hline
\textbf{Mean Queue Size} &  $K = 1$    &   $K =2$  & $K =3$ & $K = 4$ & $K = 5$\\
\hline
$m  = 1$   & 0.0408& 0.0398 & 0.0397 & 0.0397 & 0.0397 \\
\hline
$m =2$   & - & 0.0495 & 0.0495 & 0.0495& 0.0495 \\
\hline
$m =3$   & - & - & 0.0595 & 0.0595 & 0.0595\\
\hline
$m = 4$   & - & - & - &0.0696 & 0.0696\\
\hline
$m = 5$   & - & -& - &- &0.07844\\
\hline
\textbf{Mean Batch Size} &  $K = 1$    &   $K =2$  & $K =3$ & $K = 4$ & $K = 5$\\
\hline
$m  = 1$   &1.0000& 1.0009 & 1.0009 & 1.0009 &1.0009 \\
\hline
$m =2$   & - & 2.0000 & 2.0000 & 2.0000& 2.0000 \\
\hline
$m =3$   & - & - & 3.0000 & 3.0000 &3.0000\\
\hline
$m = 4$   & - & - & - &4.0000 & 4.0000\\
\hline
$m = 5$   & - & -& - &- &5.0000\\
\hline
\end{tabular}				
\end{table}

\begin{table}
\caption{Mean Queue Size and Mean Batch size for different $(m,K)$ configurations. The parameters used are $\lambda = 30$~packet/s}
\centering
\label{Table_highlambda_25ms}
\begin{tabular}{|c||c|c|c|c|c|}
\hline
\textbf{Mean Queue Size} &         $K =2$  & $K =3$ & $K = 4$ & $K = 5$\\
\hline
$m  = 1$   & 2.2972 & 1.5904 & 1.4499 & 1.4085\\
\hline
$m =2$    & 2.5720 & 1.8114 & 1.6542 &  1.6092\\
\hline
$m =3$    & - & 2.1548 & 1.9433& 1.8766\\
\hline
$m = 4$   & - & - & 2.2397 & 2.1575\\
\hline
$m = 5$   & - & - & -&  2.4345\\
\hline
\textbf{Mean Batch Size}&         $K =2$  & $K =3$ & $K = 4$ & $K=5$\\
\hline
$m  = 1$   & 1.5504 & 1.6442 &  1.6664 & 1.6710\\
\hline
$m =2$    & 2.0000 & 2.2645 & 2.3301 & 2.3468\\
\hline
$m =3$    & - & 3.0000 & 3.1455&3.1893\\
\hline
$m = 4$   & - & - & 4.0000& 4.0769\\
\hline
$m = 5$   & - & - & -& 5.0000\\
\hline
\end{tabular}				
\end{table}

\section{Conclusion}

	This paper provides a queueing model for random linear network coding scheme for time division duplexing channels with Poisson arrivals. The analysis considers that the size of the batch that is sent using random linear coding can be in a range of values, say $(m,K)$. At the time of completing service to a batch, if the queue size is below the minimum allowable value of the batch $m$, the system will wait until the queue size becomes $m$. If the queue size is in the range of batch sizes, all data packets are serviced at that time. Finally, if the queue size is greater than the maximum allowable batch size $K$, the first $K$ packets of the queue are serviced. We have used this analysis to study the mean queue size of the system and to choose the $(m,K)$ pair that minimizes it under different $\lambda$. The analysis is useful if we are interested in choosing the optimal $(m,K)$ given a different objective function.       

	 We present a complete characterization of the problem of random linear network coding for time division duplexing presented in \cite{lucaniInfocom09}, by providing a recursive expression for the moment generating function of the service time. This moment generating function is also valid for the energy to complete transmission of a batch of data packets using the appropriate substitutions.   

	Numerical results suggest that the mean queue size shows greater dependence on the value of minimum batch size $m$ than on the value of the maximum batch size $K$ for low values of $\lambda$. In general, $K$ determines the maximum serviceable $\lambda$ of the system, while $m$ should be allowed to have small values. In our examples, $m=1$ provided the best performance in terms of minimizing the mean queue size. Also, numerical results suggests that having a fixed batch size, i.e. $m=K$, is not the optimal configuration in general.   

 	Future research should consider choosing $(m,K)$ to optimize the total transmission time and total transmission energy of a packet, i.e. from the time it enters the queue until it is serviced. Also, extensions of the principles proposed for one link to the general problem of wireless networks will be studied. These extensions are possible within the framework of random linear network coding.

\section*{Acknowledgment}
This work was supported in part by the National Science Foundation under grants No. 0520075, 0831728 and CNS-0627021, by ONR MURI Grant No. N00014-07-1-0738, and subcontract \# 060786 issued by BAE Systems National Security
Solutions, Inc. and supported by the Defense Advanced Research Projects
Agency (DARPA) and the Space and Naval Warfare System Center (SPAWARSYSCEN),
San Diego under Contract No. N66001-06-C-2020 (CBMANET).




%

\end{document}